\documentclass[prd,twocolumn,floats,floatfix,nofootinbib,superscriptaddress,showpacs, textcolor]{revtex4-2}

\usepackage{graphicx}
\usepackage{dcolumn}
\usepackage{amssymb}
\usepackage{bm}
\usepackage{amsmath}
\usepackage{xcolor}
\usepackage{braket}
\usepackage{footnote}

\newcommand{\be}{\begin{equation}}
\newcommand{\en}{\end{equation}}
\newcommand{\bea}{\begin{eqnarray}}
\newcommand{\ena}{\end{eqnarray}}
\newcommand{\ie}{\textsl{i.e.~}}
\newcommand{\no}{\nonumber\\}
\usepackage{siunitx}

\begin{document}
\title{On a new prediction of Causal Quantum Mechanics for the two-slit interference experiment with electrons.}

\author{E.~Sergio Santini}
\email{santini@cbpf.br}
\affiliation{Comiss\~ao Nacional de Energia Nuclear - CNEN \\
Rua General Severiano 90, Botafogo, 22290-901,  Rio de Janeiro, Brazil}
\affiliation{Centro Brasileiro de
Pesquisas F\'{\i}sicas - CBPF - COSMO, Rua Xavier Sigaud, 150, Urca,
22290-180, Rio de Janeiro, Brazil}
\author{Germ\'{a}n M.~Chiappe}
\email{chiappegerman@yahoo.com.ar}
\affiliation{Universidad de Buenos Aires - UBA- Facultad de Farm\'{a}cia y Bioqu\'{\i}mica \\
Ciclo B\'{a}sico Com\'{u}n - CBC - \\ Jun\'{\i}n 954, C1113AAD, Ciudad de Buenos Aires, Argentina }
\author{Rafael Gonz\'{a}lez }
 \email{rgonzale@ungs.edu.ar; rgonzale@df.uba.ar}
\affiliation{
Instituto de Desarrollo Humano, Universidad Nacional de General Sarmiento\\
Pcia de Buenos Aires, Argentina}
\affiliation{
 Depto. de F\'\i sica FCEyN, Universidad de Buenos Aires, Buenos Aires, Argentina}

\date{\today}

\begin{abstract}
The causal quantum mechanics (\ie  Bohmian or de Broglie-Bohm or Bohm-de Broglie quantum mechanics) has made possible to calculate the trajectories of electrons in a typical double-slit experiment [C. Philippidis et al., Il Nuovo Cimento, 52 B, 15-28 (1979)]. The trajectories do not correspond to an uniform  movement but to an accelerated one. The acceleration is caused by the quantum potential. From the quantum theoretical point of view, there is a probability for the electron to emit photons, with a certain emission power, during its movement from the slits to the screen. 
We find a quantum general formula for the emission power of photons, valid independently of the interpretation.
Then, according to the Copenhagen interpretation, this formula gives a strictly zero value for the emission power because the electron moves as a free particle after it leaves the slit and before reach the screen. Then, there is no emission of photons.  
In the case of the causal interpretation, the emission power results, for a concrete real experiment, in a very tiny but not a zero value, driven by the square of the quantum potential gradient.
We give an idea of the type of spectrum that could be measured. A brief idea of a possible experimental arrangement in order to detect this effect, is given.
\end{abstract}



\maketitle

\section{Introduction}

In Feynman, Leighton and Sands 's words \cite{feynman3} the interference experiment with electrons "`...has in it the heart of quantum mechanics. In reality, it contains the {\it only} mystery."' 
The formation of the interference pattern has been demonstrated in several experiments, 
among them J\"{o}nsson (1961)\cite{jonsson}, Tonomura et. al (1989)\cite{tonomura}. At the same time the theoretical explanation indicated that "`The electrons arrive in lumps, like particles, and 
the probability of arrival of these lumps is distributed like the distribution of 
intensity of a wave. It is in this sense that an electron behaves sometimes like a 
particle and sometimes like a wave."\cite{feynman3}, or "`In quantum mechanics there is not such concept as the path of a particle."'\cite{landau3}. 
On the other hand the trajectories of the electrons in an typical two-slit interference experiment were computed and plotted in the framework of the causal approach to quantum mechanics \cite{phi}. This is a version of quantum mechanics that reproduces all the experimental results explained by usual quantum mechanics \cite{hol}\cite{BdB}. 


In this work we analyze the typical interference experiment of electrons for which we will show that the usual interpretation predicts that they do not radiate on their way to the screen. On the other hand we will show that the causal interpretation 
 predicts that electrons radiate with a very small power, on their way to the screen. We emphasize that this is a prediction about individual events. However, the possible detection of this radiation would be a strong and indirect experimental support in favor of Bohmian trajectories. 

When the present work was ready to be submitted to be considered for publication, we learned of a very interesting  preprint by Pisin Chen, more than twenty years ago in which a proposal similar to ours had already been discussed \cite{pis}. In that preprint an analytical model of the quantum potential of the experiment of the two slits with electrons was used and a numerical study was also carried out. It was concluded, like us in the present work, that the electrons must emit radiation.
We can say that this preprint and the present work, in a certain way, complement each other since, in our section III,  we make a graphic study in scale, somewhat handmade and in Chen's work more general methods were used. However, in addition to the fact that Chen's prediction is mainly in the visible range, a range different from that predicted in the present work, namely radio waves\footnote{However, we note that the width of the slits used in Chen's work is different.}, the same author in a more recent preprint concludes that this radiation does not really exist \cite{kpis}.Therefore, we think that our proposal is of additional interest since, using different methods, we affirm that the causal view of quantum mechanics is saying that this radiation should exist.

This letter is organized as follows:
In section II we present the typical experiment and we made the calculations following  
an elementary pedestrian approach to quantum electrodynamics, giving an exact theoretical  formula for the emission power by electrons valid independently of the interpretation. We calculate the numerical value of the emission power in the case of Copenhagen interpretation. 
In section III we make the computations according to the causal approach giving first an exact formula for the emission power by electrons and then an estimate for its value in the case of two real experimental possibilities. In section IV we say two words about polarization of the emitted radiation and 
in section V we  discuss whether the uncertainty principle would prevent the detection that radiation.
Section VI is for conclusions and discussion. A certain calculation was put in the appendix, in order not to separate the reader from the main line of reasoning.


\section{Elementary computation of the emission power of photons by an accelerated electron 
}\label{elementary}
Let's consider an usual two-slit experiment given by an electron source $S_{1}$, two slits $A$ and $B$ and a screen $S_{2}$. 

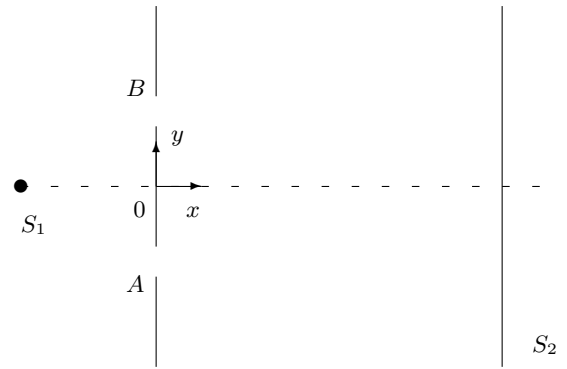
\begin{figure}
\setlength{\unitlength}{0.20mm}
\begin{picture}(400,240)
\multiput(50,120)(20,0){18}{\line(1,0){5}}
\put(140,0){\line(0,1){60}}
\put(140,80){\line(0,1){80}}
\put(140,180){\line(0,1){60}}

\put(370,0){\line(0,1){240}}
\put(140,120){\vector(0,1){30}}
\put(140,120){\vector(1,0){30}}
\put(50,120){\circle*{9}}
\put(50,90){$S_1$}
\put(390,10){$S_2$}
\put(120,50){$A$}
\put(120,180){$B$}
\put(160,100){$x$}
\put(150,150){$y$}
\put(125,100){$0$}

\end{picture}
\caption{Two-slit interference experiment for electrons.}
\label{doble}
\end{figure}

We adopt a co-ordinate system with origin at $O$ as indicated in the Fig. 
\ref{doble},  with  the centers of the slits having co-ordinates $(O, Y)$ and $(O,-Y)$. (same convention that as \cite{phi}). After go through and coming out the slits an electron becomes free, i.e. the potential acting on it is zero

 In the experiment  presented by J\"{o}nsson  in \cite{jonsson} which is the same later studied in \cite{phi}, the kinetic energy of a typical electron  is $45\,keV$. This represents approximately $9\%$ of its rest mass, $511 keV$,  and this mean a non-relativistic case \cite{phi}. Then the hamiltonian operator  of the electron is given by\footnote{In this section we follow the lines and notation of \cite{harris} Ch. 3. but keep in mind that here we have an electron instead of an atom.}.

\be\label{freee}
H_{e}=\frac{\bm{p}^2}{2 m}
\en

If we consider that a photon can be emitted, these are characterized by the potential vector $\bm{A}(\bm{r})$ and then the Hamiltonian of the total electron $ + $ photon system is

\be
H=\frac{1}{2 m}(\bm{p}-\frac{e}{c}A)^2+\frac{1}{8\pi}\int dx^3(E^2+B^2)
\en

that can be written as

\be
H= H_{e}+H_{rad}+H_{I}
\en

where

\be
H_{I}=-\frac{e}{mc}\bm{p} \cdot \bm{A}+\frac{e^2}{2mc^2}A^2
\en

\be
H_{rad}=\frac{1}{8\pi}\int dx^3(E^2+B^2) \,
\en

The term $H_{I}$ will be treated as a perturbation. The Hamiltonian without perturbation

\be
H= H_{e}+H_{rad}
\en

has as  eigenvectors

\be
\Ket{e + radiation}=\Ket{\bm{p}}\Ket{...n_{\bm{k} \sigma}..}_{rad} \, .
\en

The interaction hamiltonian $H_{I}$ induce transitions between this states, and the  transition probability per unit time is given by  the Fermi `Golden Rule'' as:

\be\label{golden}
\frac{prob}{time}=\frac{2 \pi}{\hbar}\left|M_{fi}\right|^{2}\delta(E_{f}-E_{i})
\en

where

\bea
\left|M_{fi}\right|= \Bra{f}H_{I}\Ket{i}+ \sum \frac{\Bra{f}H_{I}\Ket{n}\Bra{n}H_{I}\Ket{i}}{E_{i}-E_{f}+i\eta}+ \\ \nonumber \frac{\Bra{f}H_{I}\Ket{n}\Bra{n}H_{I}\Ket{m}\Bra{m}H_{I}\Ket{i}}{(E_{i}-E_{n}+i\eta)(E_{i}-E_{m}+i\eta)}\, +...
\ena

Now we write $H_{I}$ as

\be
H_{I}=H'+ H'' \, ,
\en

where

\be
H'\equiv -\frac{e}{mc}\bm{p} \cdot \bm{A}
\en

\be
H''\equiv \frac{e^2}{2mc^2}\bm{A}^2
\en

and substituting the expansion in normal modes for $\bm{A}$

\begin{widetext}
\begin{eqnarray*}
A(x,t)=\sum_{k, \sigma}\sqrt{\frac{2 \pi \hbar c^2}{\Omega \omega_{k}}}\bm{u}_{\bm{k} \sigma}\left[a_{\bm{k} \sigma}(t)e^{i \bm{k}\bm{r}}+a^{+}_{\bm{k} \sigma}(t)e^{-i \bm{k}\bm{r}}\right]
\end{eqnarray*}

being

\be
a_{\bm{k} \sigma}(t)=a_{\bm{k} \sigma}(0)e^{-i \omega_{k}t}
\en

\be
a^{+}_{\bm{k} \sigma}(t)=a^{+}_{\bm{k} \sigma}(0)e^{i \omega_{k}t}
\en
the destruction and creation operators, and $\Omega \equiv$ volume of the box where the electromagnetic field is quantized
\end{widetext}

\begin{widetext}
we obtain:

\begin{eqnarray*}
H'=-\frac{e}{mc}\bm{p}\sum_{\bm{k} \sigma}\sqrt{\frac{2 \pi \hbar c^2}{\Omega \omega_{k}}}\bm{u}_{\bm{k} \sigma}\left[a_{\bm{k} \sigma}e^{i \bm{k}\bm{r}}+a^{+}_{\bm{k} \sigma}e^{-i \bm{k}\bm{r}}\right]
\end{eqnarray*}

\begin{eqnarray*}
H''=
\frac{e^2}{2mc^2}\sum_{\bm{k} \sigma}\sum_{\bm{k}' \sigma'}(\frac{2 \pi \hbar c^2}{\Omega}) \frac{1}{\sqrt{\omega_{k}\omega'_{k}}}\bm{u}_{\bm{k} \sigma} \bm{u}_{\bm{k}' \sigma'} \times  \nonumber\\  \left[a_{\bm{k} \sigma}a_{\bm{k}' \sigma'}e^{i (\bm{k}+\bm{k}')\bm{r}}+ a_{\bm{k} \sigma}a^{+}_{\bm{k}' \sigma'}  e^{i (\bm{k}-\bm{k}')\bm{r}} 
+a^{+}_{\bm{k} \sigma}a_{\bm{k}' \sigma'}e^{i(- \bm{k}+\bm{k}')\bm{r}}        +a^{+}_{\bm{k} \sigma}a^{+}_{\bm{k}' \sigma'}e^{i(-\bm{k}- \bm{k}')\bm{r}}\right]
\end{eqnarray*}
\end{widetext}

To the first order in the perturbation we see that $H'$ induce transitions in which the number of photons changes in one unity (i.e $\pm 1$), since one and only one creation or destruction operator appear in each term of it.  In the same way we see that $H''$ induces changes in which two photons are emitted or two are absorbed or one is emitted and another is absorbed.

Let's consider the electron with initial state $\Ket{p_{a}}$ and final state $\Ket{p_{b}}$. We are going to analyze the emission of one photon with wave vector $\bm{k}$ and polarization $\sigma$. We write for the initial and final states:

\be
\Ket{i}=\Ket{p_{a}}\Ket{...n_{k \sigma}...}_{rad}
\en

\be\label{b}
\Ket{f}=\Ket{p_{b}}\Ket{...n_{k \sigma}+1...}_{rad}
\en

 Transitions between this states can only be induced by $H'$ in the first order contribution to $M_{fi}$.

We have

\be
\Bra{f}H'\Ket{i}=-\frac{e}{mc} \sqrt{\frac{2 \pi \hbar c^2}{\Omega \omega_{k}}}\Bra{p_{b}}\bm{p}.\bm{u}_{\bm{k}\sigma}e^{-i \bm{k}\bm{r}}\Ket{p_{a}}\sqrt{n_{\bm{k}}+1}
\en

then, from eq. (\ref{golden})

\bea\label{probt}
&& \frac{probability}{time}= \frac{4 \pi^2 e^2}{m^2 \Omega \, \omega_{k}}   (n_{\bm{k}}+1) \cdot \no &&  \left|\Bra{p_{b}}  \bm{p}.\bm{u}_{\bm{k}\sigma}e^{-i \bm{k}\bm{r}}\Ket{p_{a}}\right|^{2}\delta(E_{b}-E_{a}+\hbar \omega_{k}) ,
\ena

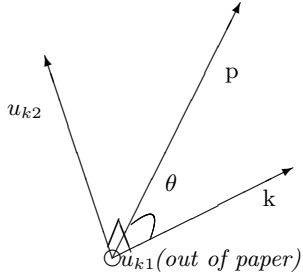
\begin{figure}
\vspace{0.5cm}
\newcounter{cms}
\setlength{\unitlength}{0.1mm}
\begin{picture}(400,400)
\put(200,0){\vector(-1,3){90}}
\put(60,190){$u_{k2}$}
\put(200,0){\vector(2,1){240}}
\put(400,70){k}
\put(200,0){\vector(1,2){170}}
\put(350,240){p}
\put(200,00){\circle{20}}
\put(210,-10){$u_{k1}$}
\put(255,-10){{\small \it(out of paper)}}
\put(194,15){\line(2,5){15}}
\put(225,9){\line(-2,5){17}}
\qbezier(225,45)(270,80)(250,25)
\put(270,90){$\theta$}
\end{picture}
\label{pola}
\caption{Polarization of photons for the sum in Eq. (\ref{sumpol}). $\bm{u_{k1}} \bot \bm{u_{k2}}$ and both perpendicular to the direction of propagation of the photon $\bm{k}$}
\end{figure}

The case $n_{k}=0$ correspond to the situation in which there is no photons a priori before the emission, in the final state. In order to calculate the lifetime of the state $\Ket{p_{a}}$ against the emission of a photon it is necessary to sum over all the possible values of $\bm{k}$ and $\sigma$ that the emitted photon can have. In summing over the polarizations we choose $\bm{u}_{\bm{k}1}$ and $\bm{u}_{\bm{k}2}$ as in Fig. \ref{pola}, then

\be\label{sumpol}
\sum_{\sigma=1,2} \left|\Bra{p_{b}}  \bm{p}.\bm{u}_{\bm{k}\sigma}e^{-i \bm{k}\bm{r}}\Ket{p_{a}}\right|^{2}=\left|\Bra{p_{b}}  \bm{p} e^{-i\bm{k}\bm{r}}\Ket{p_{a}}\right|^{2}\sin^2{\theta}
\en

Now it is a reasonable approximation to say that the emitted photon is of very low energy, i.e. it have a very long wavelength compared with the characteristic dimensions of the experiment which means $\exp{-i \bm{k}\bm{r}}\cong 1$. Substituting in (\ref{probt}), taking into account that in summing on all the possible states the probability in the numerator becomes $1$ and using the prescription

\be
\sum_{k}\longrightarrow \frac{\Omega}{(2\pi)^3}\int d^{3}k \, ,
\en
for $\Omega\rightarrow \infty$ (the volume of the box going to infinity) we obtain:

\bea\label{1sobretau}
&&\left(\frac{1}{\tau}\right)_{a\rightarrow b}=  \frac{e^2}{2 m^2 \pi} \cdot \no && \int d^{3}k \frac{1}{ \omega_{k}}  \left|\Bra{p_{b}}  \bm{p}\Ket{p_{a}}\right|^{2} \sin^2{\theta}\delta(E_{b}-E_{a}+\hbar \omega_{k})
\ena

Now we use spherical coordinates in the $k$-space with the $z$-axis in the $\Bra{p_{b}}  \bm{p}\Ket{p_{a}}$ direction, then:

\be
d^3k=k^2 dk \sin{\theta} d\theta d\phi
\en

and using $k=\frac{w_{k}}{c}$, we have

\be
d^3k=\frac{w_{k}^2}{c^3} dw_{k} \sin{\theta} d\theta d\phi
\en

which substituting in (\ref{1sobretau}) and integrating, allow us to write:

\be\label{1sobretau2}
\left(\frac{1}{\tau}\right)_{a\rightarrow b}=\frac{4e^2}{3 m^2 c^3 \hbar}\omega_{ab}\left|\Bra{p_{b}}  \bm{p}\Ket{p_{a}}\right|^{2}
\en
were $\omega_{ab}=\frac{(E_{a}-E_{b})}{\hbar}$ is the frequency of the emitted photon.

Using that $\bm{p}=m \dot{\bm{r}}$ (non relativistic case) we can re-write the bracket in (\ref{1sobretau2}) in the following form \footnote{See appendix A.}:

\be\label{31}
\Bra{p_{b}} \bm{p}\Ket{p_{a}}=\frac{i m}{-\omega_{ba}} \Bra{p_{b}}  \ddot{\bm{r}}\Ket{p_{a}}
\en

that substituted in (\ref{1sobretau2}) gives

\be\label{1sobretau3}
\left(\frac{1}{\tau}\right)_{a\rightarrow b}=\frac{4e^2 }{3 c^3 \hbar\omega_{ba} }\left|\Bra{p_{b}}  \ddot{\bm{r}}\Ket{p_{a}}\right|^{2} \, ,
\en
where we used that $\omega_{ab}=-\omega_{ba}$.

Being the energy of the emitted photon equal to $\hbar\omega_{ba}$, we have for the energy radiated per unit time (emission power)

\be\label{erut}
\left(\frac{\hbar\omega_{ba}}{\tau}\right)_{a\rightarrow b}=\frac{4e^2 }{3 c^3}\left|\Bra{p_{b}}  \ddot{\bm{r}}\Ket{p_{a}}\right|^{2}
\en

that we write as

\be\label{erut3}
\left(\frac{dE}{dt}\right)_{a\rightarrow b}=\frac{4e^2 }{3 c^3}\left|\Bra{p_{b}}  \ddot{\bm{r}}\Ket{p_{a}}\right|^{2} \, .
\en

We see a great resemblance  to the Larmor's formula for an accelerated electron of  Classical electrodynamics.

We can write the last equation using the "`fine structure constant'': $\alpha=\frac{e^2}{\hbar c}$ as

\be\label{erutbis}
\left(\frac{dE}{dt}\right)_{a\rightarrow b}=\frac{4}{3}\frac{\alpha \hbar}{c^2}\left|\Bra{p_{b}}  \ddot{\bm{r}}\Ket{p_{a}}\right|^{2} \, .
\en

It is important at this point to emphasize that the formula obtained, Eq. (\ref{erutbis}), is valid independently of the interpretation, that is, it is worth both for the interpretation of Copenhagen and for causal interpretation. \footnote{This is given in the same way that the Schroedinger's equation is valid for both Copenhagen and Causal Interpretation. The novelty will appear in the Causal interpretation when the Bohm's guidance equation is assumed (This equation, one the postulates of the Bohmian theory, reads: $ p = \nabla S, \, $ see for example \cite{hol}.) or, in other words, when it is accepted that the electron follow along trajectories, see section \ref{peca}.}

Let´s now see what the Eq. (\ref{erutbis}) says according to the Copenhagen Interpretation. 
We analyze the electron considered in our problem, which once abandoned the slits experiences no potential before striking the screen, \ie $V(x)=0$ for $x_{slit}<x<x_{screen}$. This mean that its hamiltonian is that of a free particle, Eq. (\ref{freee}), and then the operator $\ddot{\bm{r}}$ vanish:

\be\label{operace}
\ddot{\bm{r}}=\frac{i}{\hbar}[H_{e}, \dot{\bm{r}}]=\frac{i}{\hbar 2m}[\bm{p}^2, \bm{p}]=\bm{0}\, .
\en

Then, equation (\ref{erutbis})  gives:

\be\label{erut2}
\left(\frac{dE}{dt}\right)_{a\rightarrow b}=0 \, \, ,
\en

\ie a strictly null value, which means the energy emitted per unit time is zero and we have no photon emitted\footnote{In a sense we verified  the Feynman's quote \it{[...] the electron cannot emit a photon and make a transition to a different electron state while traveling along a vacuum \cite{feynmanqed}. }}. This is the answer that gives the Copenhagen Interpretation of QM. The electrons  goes from the slits to the screen without emission of photons in most of its travel, with a possible  exception in two points which are the final point in the screen and  the initial passing through the slits, points where the potential is not vanishing. Note that, following the founders Bohr, Heisenberg, and also Landau and others, we have not talked about "`trajectory"' in our deduction.

We followed an elementary computation in Q.E.D., \ie we have not made use of the 2nd quantization formalism but, as it is well known,  the answer to the problem must be the same with the only price to pay  being the lost of  manifestly covariant equations.

\section{Photon emission in the Causal approach}\label{peca}
\begin{figure}[!bh]
\centering
\includegraphics[]{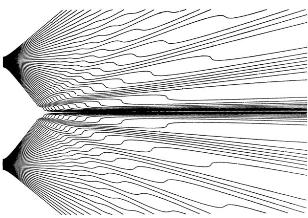}
\caption{Trajectories of the electrons in the two-slit interference experiment predicted by causal mechanics (Extracted with permission from \cite{phi}).}
\label{trayectorias}
\end{figure}

\begin{figure}[!th]
\centering
\includegraphics[]{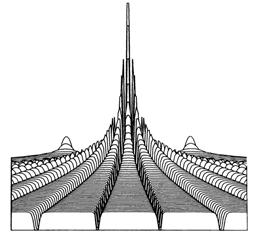}
\caption{The quantum potential for the two (gaussian) slits viewed from the screen $S_{2}$ (Extracted with permission from \cite{phi}).}
\label{qp}
\end{figure}

\begin{figure}[!th]
\centering
\includegraphics[]{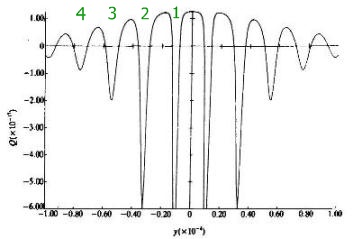}
\caption{A section of the quantum potential at 18cm from the slits $S_{2}$ (Ordinate axis: $Q [10^{-17} ergs]$, abscisa axis: $y [10^{-4}cm]$. Extracted with permission from  \cite{phi}).}
\label{qps}
\end{figure}

The two-slit interference experiment with electrons was studied in the framework of the causal quantum mechanics  in \cite{phi} where  the bohmian trajectories were first calculated (Fig. \ref{trayectorias}). An interesting discussion of this experiment at the light of causal quantum mechanics is given in \cite{hol}. The trajectory of an electron is affected by the quantum potential which is depicted in Fig. \ref{qp}. 
 The plot of the quantum potential shows, after the high spikes in the central region near the slits,  a set of troughs and plateaus. An electron emerging for one slit can be first repelled by the central spikes and then moves practically uniformly  (with a small component of velocity in the $y$-direction) until  it encounters one of the 
troughs in $Q$. One can have an idea of the variation of $Q$ in the $y$ axis by plotting the cross section which is depicted at about $18$ cm from the plane of the slits: we see a series of "`potential wells" (or valleys) corresponding to the troughs (Fig. \ref{qps}).
The electron "`fall"' in the potential well where is first accelerated with a strong force $\frac{-\partial Q}{\partial y}$ and then decelerated (the quantum potential for this experiment depends only on $y$, see \cite{phi}, \cite{hol}). From the quantum theoretical point of view we  can say that there exist a certain probability for the accelerated electron emit a photon.

The formal calculation according to the causal approach can follow along the same line section \ref{elementary}, that is to say, using theory of perturbations and therefore, it is not necessary to repeat the deduction.  The result for the 
emission power, as we already said,  is given by the same formula Eq. (\ref{erutbis}). 

However, in order to find what this equation says according to the causal interpretation, it is necessary to re-write it in terms of the quantities in which this interpretation is formulated, \ie the  properties of the particle, in this case, the spatial position of the electron. Recall that in this approach it is accepted that the electron follows trajectories. This can be done using the relation between particle properties  and quantum mechanicals operators (\cite{hol} Chap. 3.5). To do that \footnote{For an heuristic deduction see Appendix B}
we are going to re-write the bracket:

\be\label{2}
\Bra{p_{b}}  \ddot{\bm{r}}\Ket{p_{a}}
\en

by introducing two identities and, furthermore, simplifying notation by writing $\Ket{a}$ instead of $\Ket{p_{a}}$. We have:

\be\label{27}
\Bra{b}  \ddot{\bm{r}}\Ket{a} =\Bra{b} \int dr \Ket{r}\Bra{r} \ddot{\bm{r}} \int dr' \Ket{r'}\Braket{r'|a} \nonumber
\en

\be\label{4}
= \int dr \int dr' \Braket{b|r} \Bra{r} \ddot{\bm{r}}  \Ket{r'}\Braket{r'|a}
\en

Note that the acceleration operator $\ddot{\bm{r}}$ can be written in the $\Ket{r}$ representation as proportional to a delta function, \ie: $\Bra{r}\ddot{\bm{r}}\Ket{r'}= \ddot{r} \delta(r-r')$.
To see that consider the quantum dynamical equation in Heisenberg picture, which is in general

\be
m \ddot{\bm{r}}= -\frac{\partial V(\bm{r})}{\partial \bm{r}}
\en
or 
\be
\ddot{\bm{r}}= \frac{F(\bm{r})}{m}
\en
where the force operator $F(\bm{r})$ was defined as $ F(\bm{r})\equiv-\frac{\partial V(\bm{r})}{\partial \bm{r}}$ .
\vspace{0.3cm}

Then
\be
\Bra{r}\ddot{\bm{r}}\Ket{r'}= \frac{1}{m} \Bra{r}F(\bm{r})\Ket{r'}
\en

\be
\Bra{r}\ddot{\bm{r}}\Ket{r'}= \frac{F(r)}{m} \Braket{r|r'}
\en

\be
\Bra{r}\ddot{\bm{r}}\Ket{r'}= \frac{F(r)}{m} \delta(r-r')
\en 

\be\label{555}
\Bra{r}\ddot{\bm{r}}\Ket{r'}= \ddot{r} \delta(r-r')
\en 

We see that   $\ddot{r}\equiv\frac{F(r)}{m}$ is the eigenvalue of acceleration operator. It is easy to show that the local expectation value (LEV) associated with the operator acceleration  coincides with the eigenvalue of operator acceleration\footnote{See appendix C for the computation. The  LEV is  a characteristic of the particle, not of the statistical ensemble, see, for example,  \cite{hol} chapter  3.5.} $\ddot{\bm{r}}$. Substituting (\ref{555}) in  (\ref{4}) we obtain:

\be\label{6}
\Bra{b}  \ddot{\bm{r}}\Ket{a} = \int dr \Braket{b|r}\int dr' \ddot{r} \delta(r-r') \Braket{r'|a}
\en

\be\label{7}
\Bra{b}  \ddot{\bm{r}}\Ket{a} = \int dr \Braket{b|r} \ddot{r}  \Braket{r|a}
\en

The vector quantity $ \ddot{r} $, which is not an operator but a LEV, can be obtained from the dynamical equation of causal theory which, it is good to remember, comes directly from the real part of Schroedinger's equation after take the spatial derivative (the gradient). This equation reads as follows \cite{hol}\cite{BdB}:

\be\label{2law}
\frac{d}{dt} (m \dot{r})=-\nabla(V+Q)|_{r=r(t)} \, ,
\en
where $Q$ stands for the quantum potential.

In the experiment analyzed here we have $V=0$ for the electron along the trajectory from the slits to the screen (excluding the extremal points). 

Then

\be\label{2law0}
\frac{d}{dt} (m \dot{r})=-\nabla Q|_{r=r(t)} \, .
\en

or\footnote{In order not to overload the writing, from now on we implicitly understand that the gradient is evaluated in the instantaneous position of the electron $r=r (t)$.}

\be\label{acceleration}
\ddot{r}=-\frac{\nabla{Q}}{m}
\en
and substituting in  (\ref{7}) gives:

\be\label{8}
\Bra{b}  \ddot{\bm{r}}\Ket{a} = \int dr \Braket{b|r} (-\frac{\nabla{Q}}{m}) \Braket{r|a}
\en

Finally, using (\ref{8}) in (\ref{erutbis}) it is obtained

\be\label{erutbisab}
\left(\frac{dE}{dt}\right)_{a\rightarrow b}=\frac{4}{3}\frac{\alpha \hbar}{c^2 m^2}\left|\int dr \Braket{b|r} \nabla{Q} \Braket{r|a}\right|^{2}\,.
\en

This is the general answer that causal approach gives for the emission power of an individual electron in the  double slit experiment. 


It is noteworthy that we could also have considered the causal interpretation of the electromagnetic field. In this sense, it is possible to show that in such a case the formula (\ref{erutbisab}) is maintained. However, it may be of interest to give an ontological description of the radiation emission process studied, considering,  instead of the Fock states  assumed in Eq. (\ref{b}) which are equivalent to plane waves, non-stationary states. These are given by packages or superpositions of Fock states with a certain function of weight, which allow to describe in a more realistic way the process. In this case, the weight function will appear  included in the formula of the emission power. Bohm, in his 2nd article on "hidden variables" of 1952, developed the causal interpretation of the electromagnetic field and studied, in particular, the photoelectric and Compton processes from that point of view using non-stationary states \cite{BdB}. A valuable report on the causal interpretation of the electromagnetic field can be found in \cite{kaloyerou}.

\begin{figure}[!th]
\centering
\includegraphics[width=8cm]{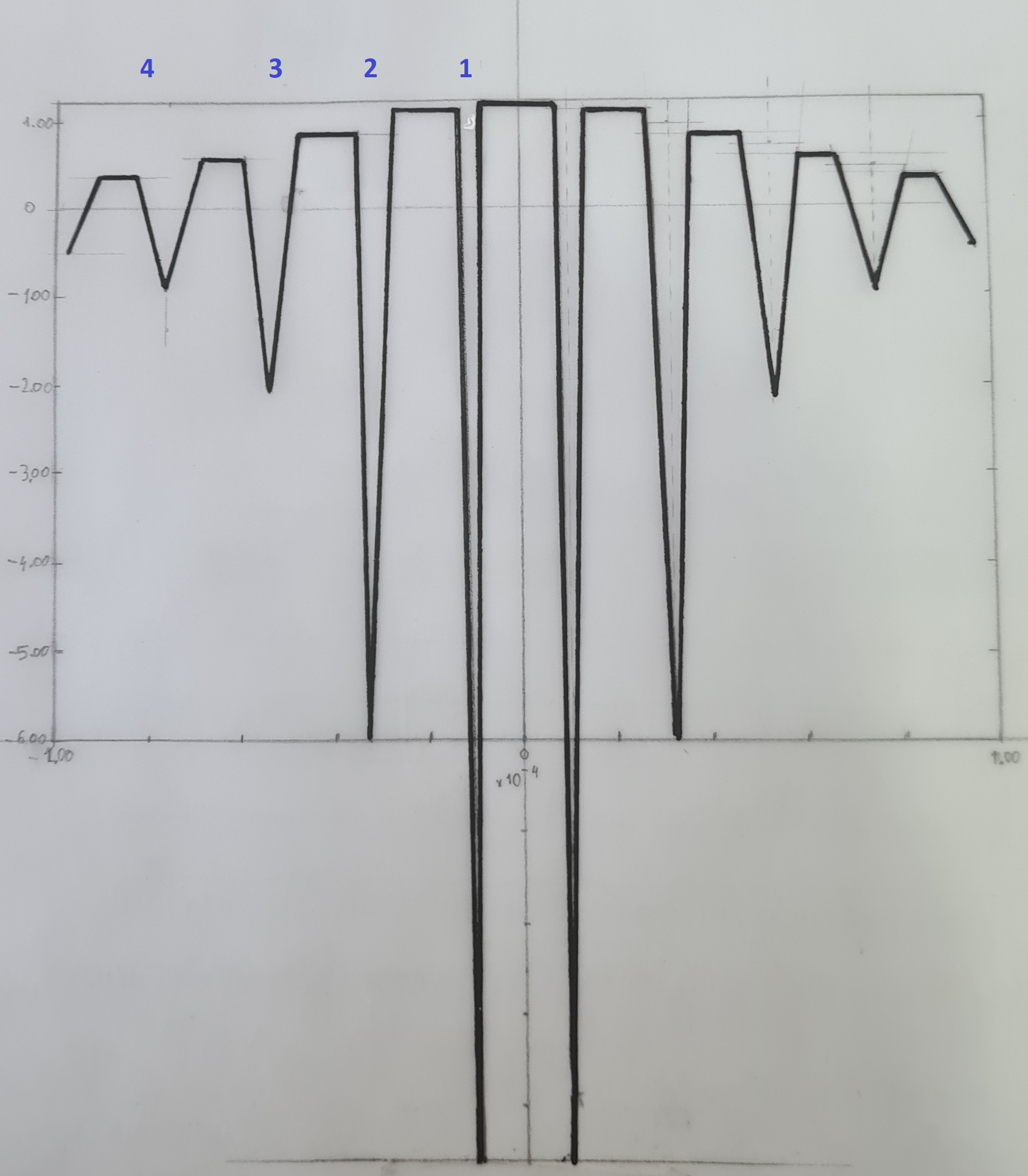}
\caption{Approximating the section of the quantum potential given in Fig. \ref{qps} (Ordinate axis: $Q [10^{-17} ergs]$, abscisa axis: $y [10^{-4}cm]$)}
\label{qptria2}
\end{figure}

In order to continue the analysis with data from a concrete and real experiment we can make a reasonable approximation:  we accept that the gradient is approximately constant or, in other words, we approximate the curvilinear walls of each well, shown in the Fig. \ref{qps}, by straight walls, as it is shown in Fig. \ref{qptria2}:

\be
\nabla Q(r)= constant= \frac{\Delta Q}{\Delta r}
\en
Then we can take it out of the integral in (\ref{erutbisab}):

\be\label{erutbis4}
\left(\frac{dE}{dt}\right)_{a\rightarrow b}=\frac{4}{3}\frac{\alpha \hbar}{c^2 m^2} (\nabla{Q})^2 \left|\int dr \Braket{b|r}  \Braket{r|a}\right|^{2}
\en
or
\be\label{erutbis5}
\left(\frac{dE}{dt}\right)_{a\rightarrow b}=\frac{4}{3}\frac{\alpha \hbar}{c^2 m^2} (\nabla{Q})^2 \left| \Braket{b|a}  \right|^{2} \, .
\en


For the square of scalar product $|\braket{b|a}|^2$, for a reasonable assumption for the initial and final  sates of the electron, it is possible to show that is practically   equal to one. For example, making the realistic assumption of  gaussians states

\be\label{gaussa}
\Braket{r|a}= (\frac{2}{\pi d^2})^{\frac{3}{4}} e^{\frac{i}{\hbar}p_{a}r-\frac{r^2}{d^2}}
\en

\be\label{gaussb}
\Braket{b|r}= (\frac{2}{\pi d^2})^{\frac{3}{4}} e^{-\frac{i}{\hbar}p_{b}r-\frac{r^2}{d^2}}
\en

we have

\be\label{escalar}
|\Braket{b|a}|^2=e^{(p_a-p_b)^2.\frac{d^2}{4 \hbar^2}}
\en
which, adopting the values:

\vspace{0.4cm}

$d=$ classical electron radius$ = 2.818 \times 10^{-13} cm$

$mc^{2}\cong0.511 \times 10^{6} eV$

$\left|p_a-p_b\right|= m\left|v_a-v_b\right| \cong m\,.\, 490923 \,cm/s$ for the valley 1 
gives

\vspace{0.4cm}
$e^{(p_a-p_b)^2.\frac{d^2}{4 \hbar^2}}= e^{3.544 \times 10^{-15}}\cong 1$

\vspace{0.4cm}

and so on, analogously for the other valleys.

So, the equation (\ref{erutbis5}) reduces very approximately  to

\be\label{erutbis11}
\left(\frac{dE}{dt}\right)_{a\rightarrow b}=\frac{4}{3}\frac{\alpha \hbar}{c^2 m^2} (\nabla{Q})^2 \, \,.
\en

Note that, using (\ref{acceleration}), this equation can be written as

\be\label{erutbis12}
\left(\frac{dE}{dt}\right)_{a\rightarrow b}=\frac{4}{3}\frac{\alpha \hbar}{c^2} (\ddot{r})^2 \, \,.
\en

It is possible estimate an approximate absolute value for the gradient $\left|\nabla Q \right|\cong \left(\frac{\Delta Q}{\Delta y}\right)$ graphically  from  Fig. \ref{qps} when the electron enters each well (recall here Q depends only on coordinate $y$: $Q=Q(y)$). The  maximum absolute value of $Q$ is, for the J\"{o}nsson experiment, approximately equal to $10^{-4}eV$ \cite{phi}\cite{hol}. From this we roughly estimate that for the 2nd well (counting from the symmetry center between the slits)  we have  a variation of $\frac{7}{16} \times 10^{-4} eV $ along a distance of $\frac{3}{70} \times 10^{-4} cm $.  Then:

\be\label{estimate}
\nabla Q\cong \frac{\Delta Q}{\Delta y}\cong \frac{\frac{7}{16}\times10^{-4}eV}{\frac{3}{70}\times10^{-4}cm}\cong 10.21 \frac{eV}{cm}
\en



Substituting (\ref{estimate}) in (\ref{erutbis11}) and using the standard values:

\vspace{0.4cm}
$\hbar\cong0.65 \times 10^{-15} eV.s$

$mc^{2}\cong0.511 \times 10^{6} eV$

$c\cong 3\times 10^{10} \frac{cm}{s}$

$\alpha\cong\frac{1}{137}$
\vspace{0.4cm}

we obtain for the mean emission power:

\be\label{erutbis3}
P_{2}\equiv\left(\frac{\Delta E}{\Delta t}\right)\cong 3.64\times 10^{-25} W \,.
\en

In the same way, for the 3rd well we have approximately $\frac{\Delta Q}{\Delta y}\cong 2.93\frac{eV}{cm}$ and the mean emission results in

\be
P_{3}\cong 2.99 \times 10^{-26} W \, .
\en

For the 4th well we obtain $\frac{\Delta Q}{\Delta y}\cong 1.17 \frac{eV}{cm}$ and the emission power is

\be
P_{4}\cong 4.78\times 10^{-27} W \,.
\en

Still for the first well, by extrapolating below the abscissa axis, we can obtain  $\frac{\Delta Q}{\Delta y}\cong 25.41  \frac{eV}{cm}$ and for the emission power

\be\label{erutbis1er}
P_{1}\cong 2.25\times 10^{-24} W \,.
\en

It is possible to have a crude idea for the frequency of some of the photons emitted.  For that we estimate the time the electron takes to cross the well (this is the {\it collision time}, call it $\tau$)\footnote{Strictly we call {\it collision time} to the time during which the electron "`feels"' the potential. That occurs in two stages, first acceleration and then deceleration, each lasting $\tau$. In each of these lapses there is a change in the velocity of absolute value $\left|\Delta v\right|$. This is the characteristic time that will define the frequency of cutoff of the spectrum, see below.}. We consider that the movement during the passage through the valley in the $y$ axis occurs approximately with  constant average accelerations (in the first half of the valley accelerated and in the second half decelerated)  which can be obtained graphically. Furthermore we make another simplification which is to consider each valley as symmetrical along its central vertical axis.

 Using for the initial $y$-velocity, in the border of the valley,  the value from the experiment performed by J\"{o}nsson, $v_{y} =1,5 \times 10^{4}\frac{cm}{s}$ we found approximately for the 2nd valley (or well) 

acceleration: 

\be
\left|a_2\right|\cong \frac{10.21}{m_e}\frac{eV}{cm}\cong  1.798\times 10^{16}\frac{cm}{s^2}
\en

collision time:

\be
\tau_{2} \cong 2.10\times10^{-11} s \, . 
\en
\vspace{0.5cm}

Accepting that during this time a photon is emitted, by using the mean emission power given by Eq.(\ref{erutbis3}) we have for the energy of the photon:

\bea
E_{2}= P_{2} \tau_{2} = 3.64\times 10^{-25} W \times  2.10\times 10^{-11} s \\ \nonumber = 7.64 \times 10^{-36}j \, .
\ena

To this photon must correspond a frequency

\vspace{0.5cm} 

\be
\nu_{2} = \frac{E_2}{h}=\frac{7.64\times 10^{-36}j}{6.63\times10^{-34}j.s}=1.15 \times 10^{-2} \, Hz \,.
\en

\vspace{0.5cm}





For the photons emitted as en electron cross  the 3rd valley we found the collision time $\tau_{3} \cong  4.42 \times 10^{-11}s $ and a frequency $\nu_3$    approximately equal to:

\vspace{0.5cm}
\be
\nu_{3}\cong 1.998 \times 10^{-3} \, Hz \,.
\en
\vspace{0.5cm}



In the same way we obtain the collision time and a frequency for the 4th and still for the 1st valley:

\vspace{0.5cm}

\be
\tau_{4} \cong  7.69 \times 10^{-11}s \, ,
\en

\be
\nu_{4} \cong 5.54 \times 10^{-4} \, Hz \, ,
\en
\vspace{0.5cm}

\be
\tau_{1} \cong  1.10 \times 10^{-11}s \, ,
\en

\be
\nu_{1} \cong 3.73 \times 10^{-2} Hz \, .
\en
\vspace{0.5cm}

Then, the electron irradiate soft photons (\ie photons with small energies compared to the energy available in the experiment) and this is a key information because the emission of soft photons by accelerated electrons was already studied in \cite{jackson3}. We can take advantage of the results obtained there in order to estimate qualitatively  the emission spectrum for all frequencies. It is important to note that the results presented by \cite{jackson3}, in particular Eq. (15.2)  "`holds quantum mechanically as well as classically"' (page 709). In the case of non-relativistic collisions there is significant radiation for when the following condition is satisfied

\be\label{condition}
\omega  < \frac{1}{\tau} \, ,
\en

where $\omega$ is the angular frequency, \ie $ \omega= 2\pi\nu$ and $\tau$ is the collision time.

\begin{figure}[!th]
\centering
\includegraphics[]{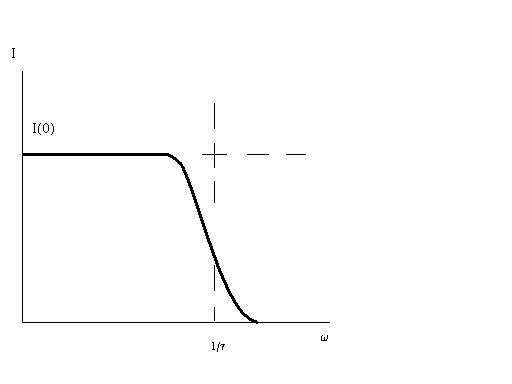}
\caption{Qualitative frequency spectrum for the emitted photons when the electron cross each well of the quantum potential. Here, I(0) is given by Eq.(\ref{erutbis9}).}
\label{step}
\end{figure}

The collision time $\tau$   is  the time the electron undergoes the acceleration in each  potential well (estimated before) and $\frac{1}{\tau}$ is, according to Eq.(\ref{condition}), the maximum angular frequency.
For $\omega  > \frac{1}{\tau}$ the  energy irradiated per unit of frequency interval (\ie the frequency spectrum) fall rapidly to zero. This spectrum will have a cutoff at that frequency and higher frequency photons will practically not be emitted \cite{jackson3}. Then the spectrum will be something like a step function with the cutoff in $\frac{1}{\tau}$, as in Fig.\ref{step} (see too \cite{jackson1} Fig. 15.1.). 

The "`height of the step"', call it $I(0)$, \ie the intensity at zero frequency, can be obtained by re-writing  Eq. (\ref{erutbis12}) in a finite form as

\be\label{erutbis7}
\delta E \delta t=\frac{4}{3}\frac{\alpha \hbar}{c^2}\left(\Delta v\right)^{2}
\en
\vspace{0.4cm}

and being the frequencies of the photons tending to zero we can write $\delta\nu\cong \frac{1}{\delta t}$, and we have in that limit:

\be\label{erutbis8}
I(0) \equiv\frac{dE}{d\nu}=\frac{4}{3}\frac{\alpha \hbar}{c^2}\left(\Delta v\right)^{2} \, ,
\en
or equivalently

\be\label{erutbis9}
I(0) \equiv\frac{dE}{d\nu}=\frac{4}{3}\frac{\alpha \hbar}{m^2 c^2}\left(\nabla Q \right)^{2} \tau^2 \,\,
\en

which represents the energy irradiated per unit of frequency at very low frequencies. 


\begin{widetext}

\begin{table}[]
\begin{tabular}{||c|c|c|c|c|c||}\hline\hline

\ Valley &$\omega_{c}=\frac{1}{\tau}$ & $\lambda_{c}$ & $I(0)$&$P_{T}$&$P_{J}$ \\ \hline

1   &$\frac{1}{1.10\times 10^{-11}s} = 9.09 \times 10^{10} Hz $ & $3.30 \, mm $  & $1.73 \times 10^{-27}\frac{eV}{Hz} $&$2.25\times 10^{-24} W$&$1.27\times 10^{-16}W$\\ \hline

2   &$\frac{1}{2.10\times 10^{-11}s} = 4.76 \times 10^{10} Hz $ & $6.32 \,mm $  & $1.03\times 10^{-27}\frac{eV}{Hz}$&$3.64\times 10^{-25} W$&$2.04\times 10^{-17}W$  \\ \hline

3   &$\frac{1}{4.42\times 10^{-11}s}= 2.26 \times 10^{10} Hz    $ & $1.33 \,cm $ & $3.35 \times 10^{-28}\frac{eV}{Hz}$&$2.99\times 10^{-26} W$&$1.68\times 10^{-18}W$   \\ \hline

4   &$\frac{1}{7.69\times 10^{-11}s}= 1.30 \times 10^{10} Hz   $ & $2.31 \,cm $ & $1.74 \times 10^{-28}\frac{eV}{Hz}$ &$4.78\times 10^{-27} W$ &$2.67\times 10^{-19}W$\\ \hline

\hline\hline 
\end{tabular}
\label{tabla}
\caption{Characteristics of the emission spectrum (Fig. \ref{step}), \ie cutoff frequency $\omega_{c}$; minimum wavelength $\lambda_{c}$;  and intensity $I(0)$  for each valley crossed by the electron, predicted according to causal quantum mechanics. The last two columns shows the emission power for two currents: $I_T$(Tonomura) and $I_J$(J\"{o}nsson) respectively, see below.}
\end{table}

\end{widetext}

So for each valley there is a spectrum as in Fig.\ref{step} each one of them with a cutoff angular frequency $\omega_{ci} \equiv \frac{1}{\tau_{i}}$ ( wavelength  $\lambda_{ci}$) and "`height of the step"' $I(0)_i$ ($i=1,2,3,4$) as indicated in the table I.

The spectrum should be given by a succession of step-type spectra (Fig.\ref{step}), one for each valley of the quantum potential that is crossed by the electron, and each one with its cutoff frequency given above and with its corresponding "height" (Eq.\ref{erutbis9}).

The results obtained correspond to the case in which a single electron emits radiation, it means a geometric arrangement like the J\"{o}nsson's experiment but in which the current is very weak. For example, a current as in the Tonomura's experiment in which it is used a current $I_{T}=1.6\times 10^{-16}A$ or $10^{3}electron/s$ arriving to the screen. Under such conditions, the average distance between successive electrons is $150$ Km and
therefore, there is practically no chance for two electrons to be present simultaneously between the slits and the screen \cite{tonomura}.
In such a way the values found here are compatible with an actual experiment \ie, J\"{o}nsson 's experiment with current $I_{T}$.

On the other hand, for the case of a greater current, say density current of electrons equal to $j=30\frac{mA}{cm^{2}}$, exactly like the one used in the J\"{o}nsson experiment, it is possible to estimate the emission power in the following way. Using that

\be\label{electron}
1.6022\times 10^{-19} C= 1e^{-}
\en

where $e^{-}$ is the electron charge, we have

\vspace{0.4cm}
$j=1.87245\times 10^{17} \frac{e^{-}}{s.cm^2}$ \, . 
\vspace{0.4cm}

And, if we consider that the total area, $S$, through which that density current flows corresponds to the area of the slits with size equal 
to: $0.3 \mu \times 50 \mu$ each one \cite{jonsson}, we have

\vspace{0.4cm}
$S= 2\times$ slit area $= 2\times 0.3\times 10^{-4}cm \times 50\times 10^{-4}cm=3\times 10^{-7}cm^{2}$,
\vspace{0.4cm}

and the total current $I_{J}$ (where J stands for J\"{o}nsson ) is

\bea
I_{J} = j . S = 1.87245\times 10^{17} \frac{e^{-}}{s.cm^2} . 3\times 10^{-7}cm^{2} \no \cong 5.6\times10^{10}\frac{e^{-}}{s}
\ena

On the other hand we can write  (\ref{erutbis3}) as

\be
P_{2}=3.64 \times 10^{-25} W = 3.64 \times 10^{-25} V. \frac{C}{s} \, ,
\en
so, using (\ref{electron}) we can write

\bea
P_{2}= 3.64 \times 10^{-25} V\times 6.24 \times 10^{18}\frac{e^{-}}{s}
\no =2.27\times 10^{-10} V \times 10^{3}\frac{e^{-}}{s}
\ena

and then

\be
P_{2}= 2.27\times 10^{-9} V \times I_{T} \, . 
\en
\ie the power is proportional to the current so, for the experiment with current $I_{J}$, we can write

\bea
P_{2J}=&&2.27\times 10^{-9} V \times I_{J}=2.27\times 10^{-9} V \times 5.6\times10^{10}\frac{e^{-}}{s} 
\no = && 2.04 \times 10^{-17} W= 2.04 \times 10^{-10}\frac{erg}{s}\, .
\ena

In the same way, from Eq.(\ref{erutbis1er}), it can be obtained for the 1st valley:

\bea
P_{1J}\cong 1.27\times 10^{-16}W = 1.27\times 10^{-9} \frac{erg}{s} \, ,
\ena

And so on for the others wells, see table I. In this way in the case of a current as J\"{o}nsson's experiment it is obtained an emission  power  several orders of magnitude greater than that obtained before for a current as Tonomura's experiment and therefore, the emitted radiation will have a greater probability of being detected in an experiment like the one described. The radiation emitted by the electron reaches the screen before it and, if it is not too attenuated, we think it could be detected by an appropriate antenna after passing through it.

To have a brief idea of how difficult it could be to measure powers as small as those obtained
 we can compare, for example, with the flux power of the cosmic microwave background radiation (CMBR): the average flux power arises from the Stefan-Boltzman law for the temperature $T = 2.73 K$ resulting in $ 3.15\times10^{-6}W/m^2 $.
In our case, using the value for the first valley $ P_ {1J} $ (J\"{o}nsson's current) and considering that the emission of radiation occurs predominantly along the valleys towards the screen, the radiation will strike on a small surface determined by the separation of the fringes (say $7000 \mathrm{\AA}$ [3]) and the height of the screen, say 10 times the separation of the fringes. This gives a flow of $\approx 2.59\times10^{-5}W/m^2$ which is only an order greater than the CMBR flux. For the radiation coming from the other valleys the flux power is even less, and even lower in the case of an experiment with Tonomura`s current.

An experimental arrangement that could eventually allow the detection of the predicted microwaves is one that amplifies them through a MASER. The region of the screen, which is where the expected microwave emission is greatest, should be able to be arranged so that they feed, along with the input signal, the Maser's resonance chamber. In this way, the predicted microwaves should produce disturbances in the output signals, normally amplified \cite{maser}.

\section{Angular distribution with a definite state of polarization}
We can say something in relation to the angular distribution of the emitted radiation and its polarization. The key in this experiment is that the acceleration imparted by the quantum potential to the electron has only a component in the "y" direction. Therefore, making use of the results presented in \cite{jackson3} it is possible to see that for the angular spectral distribution, $I$, will survive  only the contribution due to this polarization direction, \ie $I_{\bot}$, see Eq. (15.10) of that reference   (remember that they are also valid from the quantum point of view). 
We will leave for a next investigation the details of this issue.

\section{Could the detection of the predicted radiation be ruled out by the Heisenberg principle?}
Heisenberg's uncertainty principle is an inequality that is verified between two statistical dispersions, for example position and linear momentum. Since the two visions of quantum mechanics, Copenhagen and BdB coincide in all statistical predictions, this principle is also verified in the vision of BdB \cite{hol}, \cite{bh}. However, in this last view there have been different interpretations of that principle. It is not our intention to discuss this matter now and the interested reader can consult for example \cite{hol}. Here we simply want to give an idea for the answer to the question placed in the title of this section. That is why we will consider an usual application of the principle. By placing a detector beyond the screen we can accept that hypothetically emitted photons will be absorbed by the detector to produce a signal, and this will occur in a certain time interval. We might think that if the energy of the photons is too small then the detection time would be extraordinarily large, larger for example, than the age of the Universe and, in that case, that would tell us that detection would not be possible. But that is not the case, as we will see. We can make an estimate for the case of the valley that produces photons of the lowest maximum energy, namely valley 4. If we consider that in the detection process all the energy corresponding to the frequencies between zero and maximum is absorbed, this energy is given by the area below the graph (figure 7). Approaching this area by a rectangle we have that the energy is: $\Delta E \cong I(0) . \omega_c$ and for this valley, from Table I we have $\Delta E \cong 2.26 \times 10^{-18} eV $. The uncertainty relation for energy and time, $\Delta E \Delta t \geq \hbar$ gives:

\be
\Delta t \geq 288 s.
\en
which means that all those photons are detectable in about $4.8$ minutes and the answer is NO.
We have implicitly assumed that the efficiency of that detector is equal to $1$, that is, it is a perfect detector. We can instead assume, as it really is, that our detector is not perfect. For example, it has an efficiency equal to $0.5$, that is, it detects half of the photons that reach it. This will give twice the time, $9.6$ minutes and the answer is still NO. We see that, in order for the detection time to be, for example, of the order of the age of the universe (which means completely impossible detection) the efficiency should be absurdly small, which would not be real.

\section{Discussion and conclusion}
It has been demonstrated from usual quantum mechanics (interpretation of Copenhagen), that in the two-slits experiment of interference with electrons, they do not emit radiation in their way from the slits to the screen. On the other hand, but for an individual event, using
causal quantum mechanics we have shown that these electrons must emit radiation. The reason for this emission is that the quantum potential accelerates the electrons. For realistic experimental parameters compatible with experiments already carried out, we have shown that the emission spectrum can be approximated by a succession of step functions, each of them characterized by a cutoff frequency and a certain intensity. They are radio waves  with wavelengths that go approximately from  few millimeters  onwards, of very low power, but probably detectable. If the emission were detected, we would have indirect experimental evidence of the existence of the trajectories of these electrons.
Although our  numerical predictions (Table I)  was demonstrated for electrons that cross the valleys in a certain region (18 cm from the slits) the effect (Eqs. (\ref{erutbisab})(\ref{erutbis11}) ) is still  quite general and the numerical predictions can be improved.

Note that Chen's prediction \cite{pis} indicates electromagnetic waves in the visible range (using slits of another thickness) although in truth, as we observed earlier, the same author has completely refuted his own prediction \cite{kpis}.

Could these two different predictions, Copenhagen and causal quantum mechanics, be made, in some way, compatible? Let's see: we know that if the following three conditions are satisfied (\cite{BdB}):

1. $\psi$ satisfies the Schroedinger's equation.

2. $\bm{p} = \nabla S$.

3. $\rho = \left|\psi\right|^2$,

where $\rho$ is the probability density of the ensemble of particles, then all statistical predictions of the BdB interpretation coincide with the predictions of Copenhagen quantum mechanics (and these are all statistical predictions).

In fact, in the problem studied in the present work we have not made the assumption 3 because we have not needed it for our deduction: the prediction in this case has been about an individual event (the accelerated electron). This is the reason why, in principle,  there is no contradiction between the two predictions, that of Copenhagen and that of causal. The latter is a prediction about the subquantum world, which simply does not exist for the interpretation of Copenhagen. The first is a statistical prediction, like any prediction of the Copenhagen quantum mechanics \footnote{An interesting discussion on the measurement of observables in the causal theory, for the cases of individual measurements and for cases of statistical measurements, is found in \cite{bric}.}.

However, the statistical prediction in the causal interpretation continues to show that there is a no null emission of radiation, as is possible to verify by taking the expectation value in our result Eq.(\ref{erutbis11}). Writing the wave function of the electron as  $\Psi = R(r).\exp{\frac{i}{\hbar}S(r)}$, we have:

\begin{figure}[!h]
\centering
\includegraphics{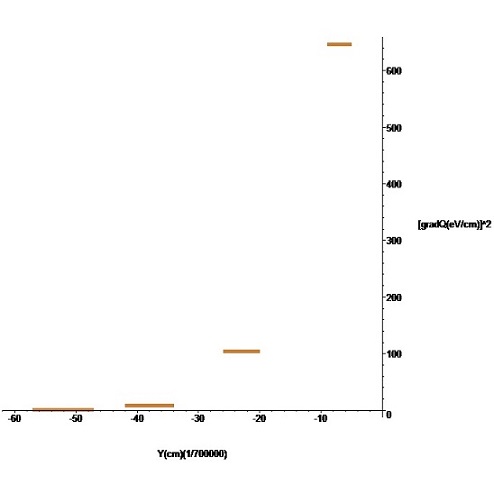}
\caption{$\nabla Q^2[\frac{eV}{cm}^2]$ vs $y[\frac{cm}{7 \times 10^5}]$, the integrand in Eq.(\ref{nose23}), coming from Fig.(\ref{qptria2}).}
\label{gradiente2}
\end{figure}

\be\label{nose22}
\left\langle \frac{dE}{dt} \right\rangle = \frac{4}{3}\frac{\alpha \hbar}{m^2 c^2}\left\langle  (\nabla Q)^2\right\rangle 
\en

\be
= \frac{4}{3}\frac{\alpha \hbar}{m^2 c^2} \int R^2 (\nabla Q)^2 d^3x  \, ,
\en

the gradient of $Q$ has only $y$ component so we have

\be \label{nose23}
\left\langle \frac{dE}{dt} \right\rangle = \frac{4}{3}\frac{\alpha \hbar}{m^2 c^2} \int R^2(y) (\frac{\partial Q}{\partial y})^2 dy  \,.
\en

In Fig.(\ref{gradiente2}) is presented an approximation to $(\frac{\partial Q}{\partial y})^2$ coming from Fig.(\ref{qptria2}).

The integrand in Eq. (\ref{nose23}) is always positive then the integral  can not be annulled\footnote{Note that in \cite{hol} Chapter 3.8.3 is demonstrated that in general $\left\langle  \nabla Q\right\rangle = 0$ for $\bm{r} \rightarrow \infty $ assuming that $R\longrightarrow 0$ at infinity, but in Eq. (\ref{nose22}) it is present the square of $\nabla Q$.} .

In fact, using the expression for $R$ given in \cite{hol} Chapter 5 formula (5.1.10) and the values for $(\frac{\partial Q}{\partial y})^2$ coming from Fig.(\ref{gradiente2}),  we obtained an approximate value for the expectation value of the square of the gradient of the quantum potential:

\be
\left\langle  (\nabla Q)^2\right\rangle \cong 164,6 (\frac{eV}{cm})^2
\en

and for Eq. (\ref{nose22}) we obtain:

\be\label{nose24}
\left\langle \frac{dE}{dt} \right\rangle \cong 3,59\times 10^{-6} \frac{eV}{s} \cong 5,74 \times 10^{-25}W \, .
\en


Therefore, the result we find is strengthened and remains to be seen how the agreement with the interpretation of Copenhagen would be redefined.

Until now, as far as we know, it has not been shown that the Bohmian trajectories can not be measured. Nor have they been measured. We affirm that the experimental detection of the predicted radiation would be an indirect test of the existence of Bohmian trajectories. In that case, we would umildly talk about a new quantum effect. In case it was not detected, the Bohm-deBroglie interpretation would be in trouble.



\vspace{0.4cm}

\section{Acknowledgements} ESS would like to thank the support of CNEN and CBPF - MCTIC - Brazil. GMCh wishes to thank the UBA - University of Buenos Aires, Argentina, for its support and RG wishes to thank UNGS-IDH and FCEyN-DF UBA, Argentina for their support. 

\section{Conflict of interest statement} On behalf of all authors, the corresponding author states that there is no conflict of interest.

\section{Appendix}

\subsection{Computation of equation (\ref{31})}

\bea
\dot{\bm{r}}=\frac{i}{\hbar}[H,\bm{r}]=\frac{i}{\hbar}[H_{e}+H_{rad},\bm{r}]= \\ \nonumber \frac{i}{\hbar}[H_{e},\bm{r}]=\frac{i}{\hbar}(H_{e}\bm{r}-\bm{r}H_{e})
\ena
then 

\bea
\Bra{p_{b}} \bm{p}\Ket{p_{a}}= m\Bra{p_{b}}  \dot{\bm{r}}\Ket{p_{a}}= \\ \nonumber m\Bra{p_{b}}  H_{e}\bm{r}-\bm{r}H_{e}\Ket{p_{a}}\frac{i}{\hbar} = \\ \nonumber m\left(E_{b}\Bra{p_{b}}\bm{r}\Ket{p_{a}}-\Bra{p_{b}}\bm{r}\Ket{p_{a}}E_{a}\right)\frac{i}{\hbar}= \\ \nonumber  \frac{im}{\hbar}\left(E_{b}-E_{a}\right)\Bra{p_{b}}\bm{r}\Ket{p_{a}} \\ \nonumber =  im\omega_{ba}\Bra{p_{b}}\bm{r}\Ket{p_{a}}
\ena
it means that

\be\label{30}
\Bra{p_{b}} \bm{p}\Ket{p_{a}}=im\omega_{ba}\Bra{p_{b}}\bm{r}\Ket{p_{a}} \,\, \ie
\en

\be\label{<p>}
\Bra{p_{b}} \dot{\bm{r}}\Ket{p_{a}}=i\omega_{ba}\Bra{p_{b}}\bm{r}\Ket{p_{a}} .
\en

Now in the same way we can compute $\Bra{p_{b}}\ddot{\bm{r}}\Ket{p_{a}}$ and obtain:

\be
\Bra{p_{b}}\ddot{\bm{r}}\Ket{p_{a}}=i \omega_{ba}\Bra{p_{b}}  \dot{\bm{r}}\Ket{p_{a}}
\en
and using (\ref{<p>}) we have

\be
\Bra{p_{b}}\ddot{\bm{r}}\Ket{p_{a}}=- \omega_{ba}^2 \Bra{p_{b}}\bm{r}\Ket{p_{a}}
\en
that together (\ref{30}) allow us write (\ref{31}).

\vspace{1cm}

\subsection{Showing  heuristically the validity of (\ref{erutbis12})}

Although we have made a rigorous deduction of equation Eq.(\ref{erutbis12}), it is possible to "`re-obtain' it  by following the elementary considerations given by Thirring in \cite{thirring} page 7 : an electron which follows an accelerated movement must emit radiation according to classical electrodynamics. But from the quantum theoretical point of view we can only say that there exist a certain probability for the accelerated electron emit a photon \footnote{We consider the single emission of a photon because it is much more likely than multiple emissions, see \cite{thirring}.}.  

If the electron  changes its  velocity $\bm{v}$ in 
$\Delta \bm{v}$ during the time interval $\Delta t$, the photon emission probability $w$ is given in essence by the Larmor formula  by

\be
w \sim \alpha \left(\frac{\Delta \bm{v}}{c}\right)^2
\en

where $\alpha $ is fine structure constant.
 
The energy  emitted by this electron is, on the average, equal to the product of probability by the energy of the 
emitted photon

\be
\Delta E \sim \alpha \left(\frac{\Delta \bm{v}}{c}\right)^2 \frac{\hbar}{\Delta t} \, ,
\en

where the frequency of the photon is of the order $\frac{1}{\Delta t}$.

Then, for the emitted power we have

\be
\frac{\Delta E}{\Delta t} \sim \alpha \frac{\hbar}{c^2} \left( \frac{\Delta \bm{v}}{\Delta t} \right)^{2}
\en

or, in infinitesimal form:

\be
\frac{dE}{dt} \sim \alpha \frac{\hbar}{c^2} \left( \frac{d\bm{v}}{dt} \right)^{2}
\en

which is in essence  Eq.(\ref{erutbis12}), except for a numerical factor.

\subsection{The local expectation value  of acceleration}
The definition of the local expectation value (LEV), ${\cal A} (x,t)$, associated to an operator $\bm{A}$, given in \cite{hol} is:
\be
{\cal A}(x,t) = Re{\frac{\Braket{\psi|r}\int d^3r'\Bra{r}\bm{A}\Ket{r'}\Braket{r'|\psi}}{\Braket{\psi|r}\Braket{r|\psi}}}\,.
 \en

For the LEV  of the operator acceleration $\ddot{\bm{r}}$, call it $ a(x,t)$, we have:

\be
a(x,t) = Re{\frac{\Braket{\psi|r}\int d^3r'\Bra{r}\ddot{\bm{r}}\Ket{r'}\Braket{r'|\psi}}{\Braket{\psi|r}\Braket{r|\psi}}}
 \en
and using that  $\Bra{r}\ddot{\bm{r}}\Ket{r'}= \ddot{r} \delta(r-r')$ (see Eq. \ref{555})

\be
a(x,t) = Re{\frac{\Braket{\psi|r}\int d^3r'\ddot{r}\delta(r-r')\Braket{r'|\psi}}{\Braket{\psi|r}\Braket{r|\psi}}}
 \en

\be
 a(x,t) = Re{\frac{\Braket{\psi|r}\ddot{r} \Braket{r|\psi}}{\Braket{\psi|r}\Braket{r|\psi}}}=\ddot{r}
\en
\ie the LEV of $\ddot{\bm{r}}$ is its eigenvalue $\ddot{r}$. QED.

\vspace{1cm}

\subsection{Discussion, somewhat heuristic, about possible situations to investigate}
1)The emission of electromagnetic radiation in the two-slit interference experiment, if any,  is for charged particles. However, when solving the trajectories in the Causal Quantum theory, only the effect of the slits and free particle is used. Then the trajectories would be the same for a proton and for a neutron, except for the correction for the difference in mass. When emitting radiation, the kinetic energy of the charged particle changes, and that modifies the trajectory, there is a reaction due to the emission of radiation that, on average, gives a force proportional to: $(2q^{2}/c^{3}) \dddot{\textbf{R}}$. This would modify the interference fringes, and it could be verified whether there is radiation emission by comparing the interference fringes for protons and neutrons, taking into account the effect of mass. It would be necessary to calculate how much the fringes change due to the effect of having another mass and how much due to the reaction by radiation emission. That would give an experimental way to see if the phenomenon occurs. It could be done with interference from neutral and ionized atoms.

2)If an electron could accelerate and not emit electromagnetic radiation, it could also do so in the hydrogen atom, and in that case there should be other solutions other than those calculated in the Holland's book  with the Causal Quantum Theory \cite{hol}. Let us remember that, according to causal quantum mechanics, in the fundamental state of the hydrogen atom, the electron is not accelerated so there is no emission and the state is stationary (see reference \cite{hol}). This argument is in favor of the fact that there must be emission in the two-slit interference experiment. If not, then it should be investigated if the quantum potential would not have an electromagnetic origin, similar to what is proposed in stochastic electrodynamics (see, for example, \cite{boyer}), in that case the effect of the electromagnetic field would already be taken into account in the quantum potential.

\end{document}